\documentclass[journal]{IEEEtran}

\usepackage{amsmath,amssymb,amsfonts}
\usepackage{algorithmic}
\usepackage{acronym}
\usepackage{graphicx}
\usepackage{textcomp}
\usepackage{xcolor}
\usepackage{balance}
\usepackage[numbers]{natbib}
\usepackage[numbers]{natbib}
\usepackage[hidelinks]{hyperref}
\hypersetup{
    colorlinks=false,
}
\usepackage{orcidlink}

\def\BibTeX{{\rm B\kern-.05em{\sc i\kern-.025em b}\kern-.08em
    T\kern-.1667em\lower.7ex\hbox{E}\kern-.125emX}}

\begin{document}

 \title{PAPR Reduction in OFDM Systems Using Neural Networks: A Case Study on the Importance of Dataset Generalization}

\newcommand{\orcidA}{\orcidlink{0009-0002-0487-0508}}
\newcommand{\orcidB}{\orcidlink{0000-0002-1905-600X}}
\newcommand{\orcidC}{\orcidlink{0000-0002-1996-7292}}

\author{Bianca S. de C. da Silva\orcidA{},
        Pedro H. C. de Souza\orcidB{},
        and Luciano L.~Mendes\orcidC{},~\IEEEmembership{Member,~IEEE}}
        
\maketitle

\maketitle

\begin{abstract}
In~\cite{11045671}, we introduced a \ac{NN} designed to reduce the \ac{PAPR} in \ac{OFDM} systems. However, the original study did not include explicit generalization tests to assess how well the \ac{NN} would perform on previously unseen data, which prevented a comprehensive evaluation of the model's robustness and applicability in diverse scenarios. To address this gap, we conducted additional generalization assessments, the results of which are presented in this case study. These results serve both to complement and to refine the original analysis reported in~\cite{11045671}. Most importantly, the overall conclusions of the initial study remain valid: the \ac{NN} is still able to reduce the \ac{PAPR} level to a desired reference value, also with a lower computational cost, confirming the effectiveness and practical applicability of the proposed method across a more generalized setting.
\end{abstract}


%
\IEEEpeerreviewmaketitle

\acrodef{5G}{Fifth Generation of Mobile Networks}
\acrodef{6G}{Sixth Generation of Mobile Networks}
\acrodef{AI}{Artificial Intelligence}
\acrodef{ABC}{Artificial Bee Colony}
\acrodef{ACE}{Active Constellation Extension}
\acrodef{Anatel}{\textit{Agência Nacional de Telecomunicações}}
\acrodef{AGA}{Adaptive Genetic Algorithm}
\acrodef{ACLR}{Adjacent Channel Leakage Ratio}
\acrodef{NN}{Neural Network}
\acrodef{AWGN}{Additive White Gaussian Noise}
\acrodef{BBO}{Biogeography-Based Optimization}
\acrodef{BER}{Bit Error Rate}
\acrodef{BPTS}{Bilayer Partial Transmit Sequence}
\acrodef{CCNF}{Constrained Clipping Noise Filtering}
\acrodef{CUPSO}{Continuous-Unconstrained Particle Swarm Optimization}
\acrodef{CP}{Cyclic Prefix}
\acrodef{CSS}{Cyclic Shift Sequence}
\acrodef{CDF}{Cumulative Distribution Function}
\acrodef{CCDF}{Complementary Cumulative Distribution Function}
\acrodef{DL}{Deep Learning}
\acrodef{DFT}{Discrete Fourier Transform}
\acrodef{DFT-s-GFDM}{Discrete Fourier Transform Spread Generalized
Frequency Division Multiplexing}
\acrodef{DSI}{Dummy Sequence Insertion}
\acrodef{DUN}{Deep Unfolding Network}
\acrodef{EGA}{Elitist Genetic Algorithm}
\acrodef{ENC}{Enhanced Nonlinear Companding}
\acrodef{FBMC}{Filter Bank Multicarrier}
\acrodef{FFT}{Fast Fourier Transform}
\acrodef{FSC}{Frequency Selective Channels}
\acrodef{FCSM}{Fully Connected Supervised Method}
\acrodef{GA}{Genetic Algorithms}
\acrodef{GFDM}{Generalized Frequency Division Multiplexing}
\acrodef{HGA}{Hybrid Genetic Algorithms}
\acrodef{ICF}{Iterative Clipping and Filtering Intercarrier Interference}
\acrodef{ICI}{Intercarrier Interference}
\acrodef{IDFT}{Inverse Discrete Fourier Transform}
\acrodef{IFFT}{Inverse Fast Fourier Transform}
\acrodef{IoT}{Internet of Things}
\acrodef{ISI}{Intersymbolic Interference}
\acrodef{ISI}{Intersymbol Interference}
\acrodef{ISM}{Industrial, Scientific and Medical}
\acrodef{LTE}{Long-Term Evolution}
\acrodef{MCSA}{Memory-less Continuous Search Algorithm}
\acrodef{MSE}{Mean Square Error}
\acrodef{MAE}{Mean Absolute Error}
\acrodef{ML}{Machine Learning}
\acrodef{MLP}{Multilayer Perceptron}
\acrodef{MIMO}{Multiple-Input Multiple-Output}

\acrodef{NOMA}{Non-Orthogonal Multiple Access}
\acrodef{OFDM}{Orthogonal Frequency Division Multiplexing}
\acrodef{OQAM}{Offset Quadrature Amplitude Modulation}
\acrodef{OOBE}{Out-of-Band Emissions}
\acrodef{PA}{Power Amplifier} 
\acrodef{PAPTR}{Peak Average Power and Time Reduction} 
\acrodef{PAPR}{Peak-to-Average Power Ratio}
\acrodef{PHY}{Physical Layer}
\acrodef{PLSA}{Pattern-Learning Search Algorithm}
\acrodef{PMF}{Probability Mass Function}
\acrodef{PTS}{Partial Transmit Sequence}
\acrodef{PSO}{Particle Swarm Optimization}
\acrodef{PSD}{Power Spectral Density}
\acrodef{PRT}{Peak Reduction Tone Radio}
\acrodef{PRnet}{PAPR-Reducing Network}

\acrodef{QAM}{Quadrature Amplitude Modulation}
\acrodef{QPSK}{Quadrature Phase Shift Keying}
\acrodef{RAN}{Radio Access Network}
\acrodef{ReLU}{Rectified Linear Unit}
\acrodef{RMS}{Root Mean Square}

\acrodef{SLM}{Selected Mapping}
\acrodef{SGD}{Stochastic Gradient Descent}

\acrodef{TI}{Tone Injection}
\acrodef{TR}{Tone Reservation}
\acrodef{TT-RDNN}{Tensor-train Residual Deep Neural Network}
\acrodef{TVWS}{TV White Space}

\acrodef{UHF}{Ultra High Frequency}

\acrodef{VHF}{Very High Frequency}

\acrodef{ZF}{Zero Forcing}


\section{Introduction}

In~\cite{11045671}, we proposed a \ac{NN}-based method to reduce the \ac{PAPR} in \ac{OFDM} systems. However, the evaluation presented in~\cite{11045671} contained a methodological flaw: the same dataset was used for both training and testing stages of the \ac{NN}. This oversight invalidated any attempt to assess the generalization performance of the proposed method. In general, generalization can be seen as a fundamental requirement, and largely determines the ability of a data-driven method to perform well on unseen data~\cite{8663155}. Without proper statistical independence between training and testing data, the evaluation may only reflect the method’s ability to memorize the training examples, rather than its capability to capture underlying patterns that can be applied to new scenarios. Consequently, the conclusions drawn in~\cite{11045671} could not reliably demonstrate the practical effectiveness of the proposed \ac{NN} method.

In this work, we address this limitation by properly dividing the dataset into distinct training and testing sets. This ensures that the proposed \ac{NN}-based approach is evaluated using unseen data, that is, samples that never appeared during the training phase, allowing a rigorous assessment of its generalization performance. This evaluation is particularly relevant in \ac{OFDM} systems, which are widely adopted in modern wireless standards due to their high spectral efficiency and robustness against multipath fading. However, \ac{OFDM} signals inherently suffer from high \ac{PAPR}, which can lead to nonlinear distortions in power amplifiers, causing spectral regrowth and performance degradation in terms of \ac{BER}. Therefore, effective \ac{PAPR} reduction techniques are essential to ensure efficient transmission, especially in frequency selective or time varying channels. In this context, evaluating the model under realistic channel conditions, such as, \ac{AWGN} or multipath Rayleigh fading is fundamental to ensure that the learned mapping remains effective beyond the training environment. A model that overfits the training data may show promising results under static or idealized conditions but fail to maintain its \ac{PAPR} reduction capability in practical communication scenarios. Consequently, proper generalization assessment becomes a key step toward validating the applicability of data-driven solutions in real world \ac{OFDM} systems.

By performing evaluations with the proper methodology, we are now able to rigorously quantify the generalization ability of the trained \ac{NN}. The new results not only confirm the effectiveness of our approach but also provide a more comprehensive and trustworthy evaluation of the model’s performance. In this way, this work complements and strengthens the findings of~\cite{11045671}, demonstrating the importance of proper dataset handling and the critical role of generalization in data-driven methods for communications systems.

The remainder of this work is organized as follows. Section~\ref{sec:PAPR} discusses the relevance of \ac{PAPR} reduction in \ac{OFDM} systems. Section~\ref{sec:corrections} addresses the importance of evaluating the generalization performance of the proposed \ac{NN} and presents the updated results for a more comprehensive performance analysis. Finally, Section~\ref{sec:conclusion} provides the conclusion to this work.

\section{PAPR Reduction in OFDM Systems}\label{sec:PAPR}

In the time domain, the \ac{PAPR} quantifies how much the peak power of the \ac{OFDM} signal $\mathbf{x}$ exceeds its average power level~\cite{6491517}.

\begin{equation} \label{eq1}
    \mathcal{P} = \displaystyle\frac{\max \left( |\mathbf{x}|^2 \right)}{\mathbb{E} \left[ |\mathbf{x}|^2 \right]}.
\end{equation}
where $\mathbb{E}[\cdot]$ denotes the expected value operator.

One of the main drawbacks of \ac{OFDM} systems is the high \ac{PAPR} of the transmitted signal. This occurs because the \ac{OFDM} waveform is generated by the superposition of multiple independently modulated subcarriers in the time domain. When these subcarriers are added constructively, the signal can exhibit large instantaneous peaks, leading to power fluctuations where the peak amplitude significantly exceeds the average transmitted power~\cite{5286238}. Such variations degrade the efficiency of power amplifiers, which must operate with a considerable back-off to maintain linearity and avoid signal distortion. Consequently, high \ac{PAPR} not only reduces power efficiency but also increases hardware complexity and implementation costs, making it one of the major challenges in practical \ac{OFDM}-based communication systems.

\begin{figure}[ht]
    \centering
    \includegraphics[width=0.9\linewidth]{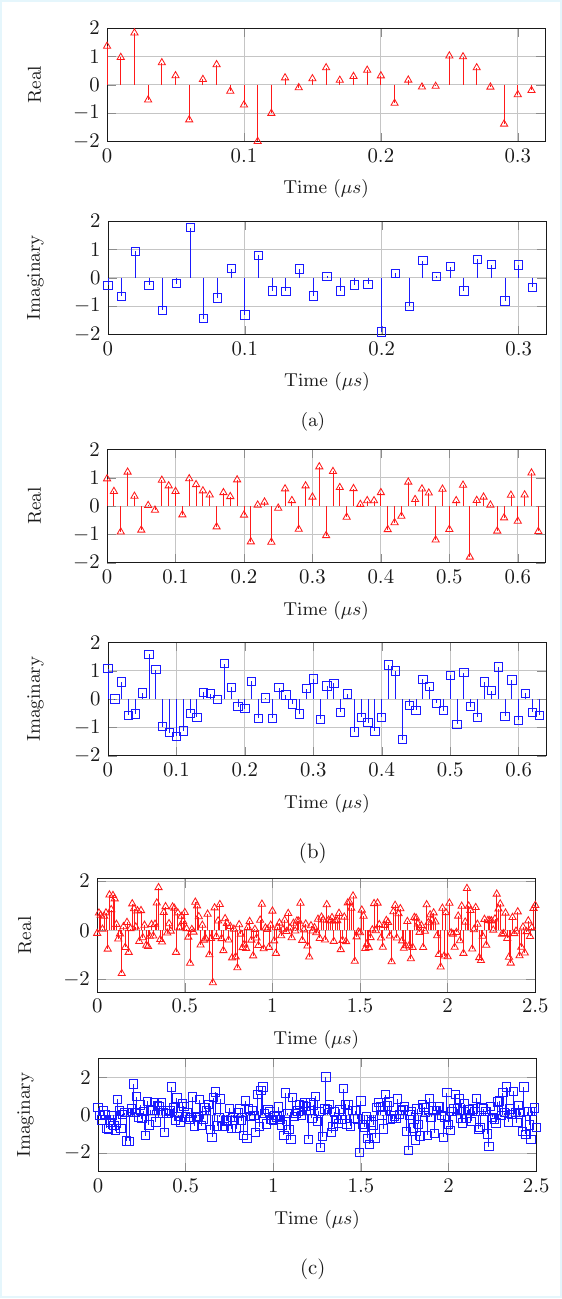}
    \caption{Illustration of \ac{OFDM} symbols in the time domain~\cite{silva2025}. (a) OFDM symbol with $K = 32$ subcarriers and $\mathcal{P} = 10.52$~dB. (b) OFDM symbol with $K = 64$ subcarriers and $\mathcal{P} = 11$~dB. (c) OFDM symbol with $K = 256$ subcarriers and $\mathcal{P} = 11.79$~dB.}
    \label{fig:papr_example}
\end{figure}

As illustrated in Fig.~\ref{fig:papr_example}, the \ac{OFDM} signal shows noticeable peaks in its instantaneous power, highlighting the presence of a high \ac{PAPR}. It is important to note that this figure was taken directly from the~\cite{silva2025} and has not been altered in any way. In this example, the waveform corresponds to an OFDM symbol in which each subcarrier is modulated with a 16-\ac{QAM} symbol, also considering various subcarrier configurations, where the number of subcarriers is denoted by $K = 32$, $64$, and $256$ and a sampling interval of $0.01~\mu\text{s}$. It can be observed that signals with a larger number of subcarriers exhibit a greater likelihood of constructive interference among them, which results in higher amplitude peaks in the time domain. Consequently, the probability of large instantaneous power values rises, leading to an increase in the overall PAPR of the transmitted signal.

To mitigate these effects, transmitters typically employ an input back-off to ensure that the \ac{PA} operates within its linear region, thereby preventing nonlinear distortion and spectral regrowth. However, this approach significantly reduces the power efficiency of the amplifier, as a considerable portion of the available power is not utilized effectively~\cite{7553476}. The resulting inefficiency leads to increased energy consumption, greater heat dissipation, and higher operational costs. These factors are particularly detrimental in portable devices, battery-powered equipment, and energy-constrained communication networks such as \ac{IoT} and mobile systems. Therefore, reducing the \ac{PAPR} of \ac{OFDM} signals is essential to allow the deployment of high-efficiency nonlinear \acp{PA} while maintaining acceptable levels of linearity, minimizing distortion, and ensuring compliance with spectral emission standards.

This issue becomes even more critical in the context of next-generation wireless technologies, such as the \ac{5G} and \ac{6G}, which employ advanced transmission techniques including dense carrier aggregation, high-order modulation formats, and massive \ac{MIMO} architectures~\cite{da2024survey}. The simultaneous use of a large number of subcarriers and spatially multiplexed antennas results in signals with a much wider dynamic range and increased amplitude fluctuations. Consequently, the likelihood of severe constructive interference among subcarriers rises, exacerbating the \ac{PAPR} problem. Furthermore, these systems often operate under stringent energy efficiency and spectral mask requirements, where nonlinear distortion can lead to adjacent-channel interference and degraded system performance. Therefore, effective \ac{PAPR} reduction techniques are indispensable to sustain the high data rates, reliability, and energy efficiency demanded by emerging \ac{5G}/\ac{6G} communication infrastructures.

In parallel with these technical challenges, the growing demand for broadband connectivity has prompted regulatory agencies worldwide to optimize the spectrum utilization. In Brazil, for instance, the \ac{Anatel}~\cite{anatel_resolucao_2021} has authorized the secondary use of underutilized frequency bands, known as \ac{TVWS}, encompassing idle television channels in the \ac{VHF} and \ac{UHF} bands. This initiative enables the deployment of Internet services in undeserved regions while maintaining protection for primary users against harmful interference~\cite{7899245}. Such regulatory frameworks not only alleviate spectrum scarcity but also promote digital inclusion and the efficient use of existing infrastructure. In this context, optimizing the power efficiency of \ac{OFDM}-based systems,  through effective \ac{PAPR} reduction techniques, is vital to guarantee reliable communication and maximize the benefits of shared and dynamic spectrum access environments.

To develop robust \ac{PAPR} reduction strategies, it is crucial to ensure that models and algorithms are trained on datasets that generalize well across different channel conditions and system configurations. The next section discusses the importance of dataset generalization and its impact on the performance and reliability of \ac{OFDM}-based systems.

\section{The Importance of Dataset Generalization}\label{sec:corrections}

Dataset generalization is one of the most important aspects of machine learning and \acp{NN} models, as it describes the model's ability to learn underlying patterns from a training dataset and apply them effectively to new, previously unseen data~\cite{5540299}. In principle, a model with good generalization ability can recognize structures and relationships embedded in the dataset that hold true not only for the training examples, but also for the overall problem domain. This ability is what distinguishes true learning from simple data memorization~\cite{273950}.

For \acp{NN}, such generalization is directly linked to how the model balances bias and variance. Bias represents the error introduced by simplifying assumptions made by the model, while variance measures how sensitive the model is to small variations in the input data. A model with high bias tends to underfit the data, failing to learn its relevant relationships. A model with high variance tends to overfit, reproducing noise and specific characteristics of the training dataset. The goal, therefore, is to achieve a balance that minimizes both errors, resulting in a network capable of representing the essential behavior of the data without becoming overly dependent on it~\cite{6708129}.

The importance of generalization becomes evident when we consider the use of \ac{NN} in practical applications, in which the data set available during training is only a limited sample of the real environment. In communication systems, for example, especially in \ac{OFDM} systems, good generalization capability is essential to ensure that the model maintains its performance under different noise, channel, or modulation conditions. A model that generalizes well is capable of adjusting parameters efficiently, such as pilot carriers or phases, even when channel characteristics change, which is essential for applications in modern and heterogeneous wireless networks~\cite{10156818}.

Several factors directly influence the generalization capability of \acp{NN}. Among these, the quantity and diversity of the dataset stand out, since larger and more varied datasets allow the model to learn more robust representations. Another crucial factor is the use of regularization techniques, such as L1 and L2 penalties, dropout, and early stopping, which limit the network's complexity and reduce the tendency to overfit~\cite{788640, 7010967}. Furthermore, the size of the architecture and the number of trainable parameters must be carefully adjusted so that the \ac{NN} has sufficient capacity to learn important relationships without becoming overly complex. The use of cross-validation and error monitoring on a validation set are also essential strategies for assessing and improving generalization~\cite{6961064}.

\subsection{Brief Review of the NN-Based PAPR Reduction Method}\label{subsec:NNmethod}

\subsubsection{Summary of the MCSA Method}

\ac{MCSA}~\cite{10639125} is a brute-force method aimed at reducing the \ac{PAPR} in \ac{OFDM} signals, based on the selective selection of pilot subcarriers. Instead of employing intelligent optimization techniques, the algorithm generates several possible pilot modifications, and the resulting \ac{PAPR} for each configuration is then evaluated.

For each \ac{OFDM} symbol $\mathbf{s} \in \mathbb{C}^K$, $N_p$ pilots are defined with values $\pm\sqrt{E}$, where $E$ is the average energy of the symbol vector $\mathbf{s}$. The corresponding \ac{OFDM} signal is then compared to a target \ac{PAPR} level of $\mathcal{P}_\text{max}$. If the \ac{PAPR} is below this threshold, the signal is accepted; otherwise, a new set of pilots is randomly selected until the target is reached or the maximum number of trials $N_t$ is reached. The signal associated with the lowest \ac{PAPR} is then transmitted.

The \ac{MCSA} requires no modifications to the receiver and can be easily scaled to a larger number of pilot subcarriers. However, its efficiency depends on the target \ac{PAPR} value: higher targets are reached quickly, while very low targets require a greater number of iterations, which increases the computational complexity of the system.

\subsubsection{The NN Method}
The architecture of the \ac{NN} is preserved identical to the one originally proposed in~\cite{11045671}, comprising a single hidden layer with 500 neurons. The loss function adopted was the \ac{MSE}, defined as the average of the squared differences between the predicted and target values~\cite{9167435}. This metric is one of the most widely used in regression tasks because it emphasizes larger deviations, thereby encouraging the network to minimize significant errors during training. In addition, the \ac{MSE} provides a smooth and differentiable objective surface, which contributes to stable convergence and consistent gradient propagation across the layers of the model.

\subsubsection{Computational Complexity Comparison}

After training, the main advantage of the \ac{NN} over the \ac{MCSA} method lies in its fixed and reduced computational complexity. While the \ac{MCSA} relies on a variable number of iterations required to find a set of pilots that meets the target \ac{PAPR} value, the \ac{NN} performs only one \ac{IFFT} during inference, keeping the complexity at ${\mathcal{O}(K^{2} + K\log_{2}K)}$~\cite{11045671}.

On the other hand, the \ac{MCSA} presents a complexity proportional to the average number of trials $v$, resulting in $\mathcal{O}(vK^{2}\log_{2}K)$, which makes it significantly more costly when the target \ac{PAPR} level is low. Thus, after training, the \ac{NN} offers a significant reduction in computational load, while maintaining comparable performance, as discussed in detail in~\cite{11045671}.

\subsection{Updated Numerical Results}

To rigorously verify the generalization performance of the proposed \ac{NN} of~\cite{11045671}, we conducted new simulations using the following \ac{OFDM} system configuration: $15$ subcarriers are modulated by \ac{QPSK} symbols, including a pair of pilot subcarriers. This setup results in a total of \(4^{15} = 1,073,741,824\) possible combinations of \ac{OFDM} symbols in the discrete sample space. While this represents only a fraction of the total possible signal combinations in larger systems, it is sufficient to test the generalization capabilities of the \ac{NN} without incurring in prohibitive computational costs. This configuration was adopted solely to limit simulation time and computational load, without compromising the validity of the generalization analysis. More complex configurations, such as of~\cite{11045671}, would require exponentially larger datasets to cover the sample space adequately, significantly increasing simulation time and computational resources. The chosen approach allows us to evaluate the model under controlled conditions, ensuring that the results reflect the model's capability to generalize rather than memorizing specific patterns of \ac{OFDM} symbols. Table \ref{tab:hip} presents a comparative summary between the parameters used here and those used in~\cite{11045671}.

\begin{table}[ht]
    \centering
    \caption{Updated Parameters}
    \begin{tabular}{ccc}
        \hline\hline
        \textbf{Parameter} & \textbf{Updated Values}  & \textbf{Previous Values}\\
        \hline\hline
        \textbf{Dataset Samples} & $2\times10^5$ and $10^6$ & $4,500$\\
        \textbf{Modulation} & \ac{QPSK} & $16$-\ac{QAM} \\
        \textbf{Subcarriers} & $15$ and $30$ & 64\\
        \textbf{Pilots} & $2$ & 2 \\
        \hline\hline        
    \end{tabular}
    \label{tab:hip}
\end{table}

Therefore, a dataset containing $2\times10^5$ randomly generated \ac{OFDM} symbols is employed, with a portion of $70\%$ allocated for training and $30\%$ for testing. In contrast with~\cite{11045671}, where the same data were inadvertently used for both training and evaluation\footnote{The normalization of the pilot subcarriers was also neglected in~\cite{11045671}. This inconsistency has been now resolved.}, this proper separation allows a valid and unbiased assessment of the model's generalization capability. Note that the training data covers less than $0.013\%$ of the total sample space, highlighting the difficulty of the task and demonstrating that successful \ac{PAPR} reduction requires the \ac{NN} to learn meaningful patterns rather than memorizing inputs.

\begin{figure}[ht]
    \centering
    \includegraphics[width=0.9\linewidth]{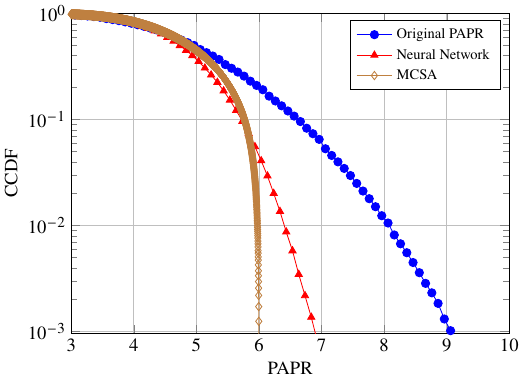}
    \caption{CCDF of the \ac{PAPR} for an \ac{OFDM} system with \ac{QPSK} modulation and 15 subcarriers.}
    \label{fig:ccdf_generalization}
\end{figure}

\begin{figure}[ht]
    \centering
    \includegraphics[width=0.9\linewidth]{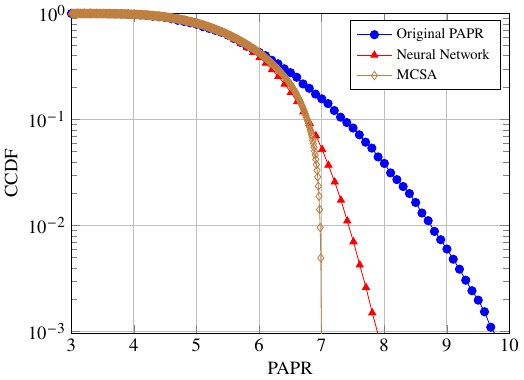}
    \caption{CCDF of the \ac{PAPR} for an \ac{OFDM} system with \ac{QPSK} modulation and 30 subcarriers.}
    \label{fig:ccdf_generalization_30}
\end{figure}

Fig.~\ref{fig:ccdf_generalization} shows the \ac{CCDF} of the \ac{PAPR} for the original signal (no \ac{PAPR} reduction), the \ac{MCSA} method~\cite{10639125}, and the \ac{NN}; considering the \ac{OFDM} system with $15$ subcarriers. The \ac{NN} achieves a substantial reduction in the \ac{PAPR} and closely adheres to the \ac{MCSA} performance, indicating that it generalizes well to unseen \ac{OFDM} symbols. Notably, for very high \ac{CCDF} values, the \ac{NN} slightly outperforms the \ac{MCSA}, which theoretically should not occur since the \ac{NN} training is supervised and uses the pilot subcarriers defined by the \ac{MCSA} as training labels. This apparent discrepancy suggests minor estimation imprecision due to the limited number of subcarriers in this configuration. Meaning that with few subcarriers (such as $15$ in this case), each subcarrier has a greater weight in determining the overall \ac{PAPR} of the symbol. Therefore, small fluctuations in the adjustment of the pilot subcarriers by the \ac{NN} can lead to noticeable variations in the \ac{CCDF}, especially at extreme values of \ac{PAPR} (\ac{CCDF} $\sim 1$), where there are fewer samples. To further investigate this discrepancy, additional simulations were performed with $30$ subcarriers, using a dataset of $10^6$ randomly generated \ac{OFDM} symbols and assuming a pair of pilot subcarriers. As illustrated in Fig.~\ref{fig:ccdf_generalization_30}, the \ac{CCDF} estimate becomes more precise with an increased number of subcarriers, and the previously observed discrepancy nearly disappears. This improvement occurs because a larger number of subcarriers effectively increases the diversity of samples contributing to the \ac{PAPR} calculation. With only a few subcarriers, extreme \ac{PAPR} values are less reliably estimated, which can lead to small apparent deviations in the \ac{CCDF}. By increasing the number of subcarriers, the sample space better represents the statistical behavior of \ac{OFDM} symbols, reducing numerical fluctuations. This observation underscores the importance of using sufficient subcarriers when evaluating \ac{CCDF} to avoid misinterpreting the performance of \ac{PAPR} reduction methods. We can also conclude from Figs.~\ref{fig:ccdf_generalization} and~\ref{fig:ccdf_generalization_30}, that increasing the number of subcarriers makes the \ac{PAPR} reduction more challenging. For example, with a pair pilot subcarriers, the \ac{MCSA} method achieves a reduction of up to $7$~dB for $30$ subcarriers, while for $15$ subcarriers it achieves up to $6$~dB.

In conclusion, these findings demonstrate that the \ac{NN} generalizes successfully to unobserved data, reinforcing the need for proper dataset separation in evaluations involving data-driven methods. Moreover, these experiments highlight the critical role of system parameters, such as the number of subcarriers and pilots, in both model training and performance assessment. Despite a small loss in spectral efficiency due to the use of pilot subcarriers, the \ac{PAPR} reduction remains substantial, demonstrating the practical relevance of \ac{NN}-based \ac{PAPR} reduction. If higher \ac{PAPR} reductions are desired, then increasing the number of pilots in systems with many subcarriers provides a viable strategy, by balancing spectral efficiency and \ac{PAPR} performance~\cite{11045671}. Therefore, properly accounting for these factors ensures that \ac{NN}-based \ac{PAPR} reduction methods can be reliably applied to practical \ac{OFDM} systems.

\subsubsection{Training Loss and Validation}

\begin{figure}[ht]
    \centering
    \includegraphics[width=0.9\linewidth]{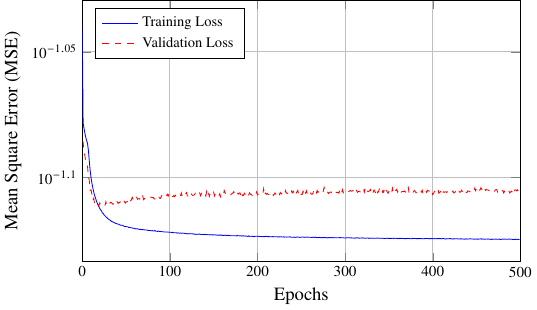}
    \caption{Training and validation loss curves using the \ac{MSE} loss over 500 epochs.}
    \label{fig:training_loss}
\end{figure}

To better understand the learning behavior of the \ac{NN} and assess its generalization capability, we first analyze the evolution of the training and validation losses. Fig.~\ref{fig:training_loss} presents the evolution of the training and validation losses over $500$ epochs. As observed, the training loss decreases steadily, and stabilizes after approximately $300$ epochs, which indicates that the learning process was performed effectively. In contrast, the validation loss stabilizes at an earlier stage and oscillates slightly around a constant value. This behavior suggests that the model quickly adapts to the validation data distribution and maintains a stable performance as the training progresses. Although a visible gap exists between the training and validation curves, this gap does not expand significantly over the epochs, implying that the \ac{NN} is capable of generalize over the dataset and of not exhibiting severe overfitting. This small gap can be explained by the intrinsic variability of the validation set, that is, the natural differences between the \ac{OFDM} symbols in the validation set and those seen during training, or by minor mismatches between the data distributions of the training and validation sets. This is a typical behavior in deep learning models trained with complex signals such as those in \ac{OFDM} systems.

As observed in Section~\ref{subsec:NNmethod}, recall that all components of the \ac{NN}, including the architecture, activation functions, loss function, and optimizer configuration, were preserved exactly as in the original model described in~\cite{11045671}, ensuring that the comparison between simulations remained fair and consistent.

\subsubsection{Code and Datasets}

The source code developed for this study is publicly available in our GitHub repository\footnote{\url{https://github.com/BiaSabrina/PAPR_Reduction_Code.git}}. This repository contains all the scripts required to reproduce the simulation results, providing a complete and reproducible framework. In addition, the dataset used in the simulations has been updated and it is also available to the interested reader\footnote{\url{https://github.com/BiaSabrina/PAPR_Reduction_NN_dataset}}. This second repository includes the dataset employed for training and testing the supervised \ac{NN} presented in Section~\ref{subsec:NNmethod}.

\section{Conclusion}\label{sec:conclusion}

In this work, we addressed two methodological inconsistencies present in~\cite{11045671}: (i) the use of identical datasets for both training and testing, which invalidated the assessment of the proposed \ac{NN} generalization capabilities, and (ii) the incorrect normalization of pilot subcarriers to unit energy. By correcting them we were able to update the numerical results and obtain a more rigorous evaluation of the \ac{NN}-based \ac{PAPR} reduction method.

The new simulations, conducted with properly separated training and testing datasets and corrected pilot normalization, confirm that the \ac{NN} generalizes effectively to unseen data. The results show that the \ac{NN} consistently achieves significant \ac{PAPR} reduction with different values for the total number of subcarriers, closely matching the performance of the \ac{MCSA} method~\cite{10639125}. Additionally, the slight variations observed in high \ac{CCDF} regions were shown to be due to a low number of subcarriers rather than a limitation of the simulation setting.

Overall, the updated numerical results presented in this work do not change the main conclusions of~\cite{11045671}: the \ac{NN}-based method can effectively reduce the \ac{PAPR} to levels close to the desired target, whilst presenting a lower computational complexity. Therefore, the proposed approach remains a valid and efficient strategy for \ac{PAPR} reduction in \ac{OFDM} systems, offering a favorable trade-off between computational complexity, spectral efficiency, and peak power constraints.

\section{Acknowledgment}\label{ch:Acknowledgment}

This work has received partial funding from the project XGM-AFCCT-2024-2-15-1, supported by xGMobile – EMBRAPII - Inatel Competence Center on 5G and 6G Networks, with financial resources from the PPI IoT/Manufacturing 4.0 program of MCTI grant number 052/2023, signed with EMBRAPII. Additionally, this work was partially supported by the Ciência por Elas project (APQ-04523-23 funded by Fapemig), the SEMEAR project (22/09319-9 funded by FAPESP), the Brasil 6G project (01245.010604/2020-14 funded by RNP and MCTI), and CNPq-Brasil.

\bibliographystyle{unsrtnat}
\bibliography{References}

@ARTICLE{11045671,
  author={S. de C. da Silva, Bianca and de Souza, Pedro H. C. and Mendes, Luciano L.},
  journal={IEEE Latin America Transactions}, 
  title={{PAPR} Reduction Technique for Mobile Communication Systems Using Neural Networks}, 
  year={2025},
  volume={23},
  number={7},
  pages={556-564},
  doi={10.1109/TLA.2025.11045671}}

@ARTICLE{5286238,
  author={Zhou, Yang and Jiang, Tao},
  journal={IEEE Transactions on Broadcasting}, 
  title={{A Novel Multi-Points Square Mapping Combined With PTS to Reduce PAPR of OFDM Signals Without Side Information}}, 
  year={2009},
  volume={55},
  number={4},
  pages={831-835},
  doi={10.1109/TBC.2009.2031465}}

@ARTICLE{7553476,
  author={Wang, Sen-Hung and Lin, Wei-Lun and Huang, Bo-Rong and Li, Chih-Peng},
  journal={IEEE Communications Letters}, 
  title={{PAPR Reduction in OFDM Systems Using Active Constellation Extension and Subcarrier Grouping Techniques}}, 
  year={2016},
  volume={20},
  number={12},
  pages={2378-2381},
  doi={10.1109/LCOMM.2016.2603529}}

@ARTICLE{788640,
  author={Vapnik, V.N.},
  journal={IEEE Transactions on Neural Networks}, 
  title={An overview of statistical learning theory}, 
  year={1999},
  volume={10},
  number={5},
  pages={988-999},
  doi={10.1109/72.788640}}

@article{da2024survey,
  title={A Survey of PAPR Techniques Based on Machine Learning},
  author={da Silva, Bianca S de C and Souto, Victoria DP and Souza, Richard D and Mendes, Luciano L},
  journal={Sensors},
  volume={24},
  number={6},
  pages={1918},
  year={2024},
  publisher={MDPI},
  doi={10.3390/s24061918}
}

@ARTICLE{5540299,
  author={Hou, Muzhou and Han, Xuli},
  journal={IEEE Transactions on Neural Networks}, 
  title={Constructive Approximation to Multivariate Function by Decay RBF Neural Network}, 
  year={2010},
  volume={21},
  number={9},
  pages={1517-1523},
  doi={10.1109/TNN.2010.2055888}}

@INPROCEEDINGS{6491517,
  author={Hsu, Chau-Yun and Liao, Hsuan-Chun},
  booktitle={2012 IEEE 11th International Conference on Signal Processing}, 
  title={PAPR reduction using the combination of precoding and Mu-Law companding techniques for OFDM systems}, 
  year={2012},
  volume={1},
  number={},
  pages={1-4},
  doi={10.1109/ICoSP.2012.6491517}}

@INPROCEEDINGS{8663155,
  author={Saravanan, R. and Sujatha, Pothula},
  booktitle={2018 Second International Conference on Intelligent Computing and Control Systems (ICICCS)}, 
  title={A State of Art Techniques on Machine Learning Algorithms: A Perspective of Supervised Learning Approaches in Data Classification}, 
  year={2018},
  volume={},
  number={},
  pages={945-949},
  doi={10.1109/ICCONS.2018.8663155}}

@ARTICLE{10156818,
  author={Merluzzi, Mattia and Borsos, Tamás and Rajatheva, Nandana and Benczúr, András A. and Farhadi, Hamed and Yassine, Taha and Müeck, Markus Dominik and Barmpounakis, Sokratis and Strinati, Emilio Calvanese and Dampahalage, Dilin and Demestichas, Panagiotis and Ducange, Pietro and Filippou, Miltiadis C. and Baltar, Leonardo Gomes and Haraldson, Johan and Karaçay, Leyli and Korpi, Dani and Lamprousi, Vasiliki and Marcelloni, Francesco and Mohammadi, Jafar and Rajapaksha, Nuwanthika and Renda, Alessandro and Uusitalo, Mikko A.},
  journal={IEEE Access}, 
  title={{The Hexa-X Project Vision on Artificial Intelligence and Machine Learning-Driven Communication and Computation Co-Design for 6G}}, 
  year={2023},
  volume={11},
  number={},
  pages={65620-65648},
  doi={10.1109/ACCESS.2023.3287939}}

@ARTICLE{6708129,
  author={Sohn, Insoo},
  journal={IEEE Communications Letters}, 
  title={A Low Complexity PAPR Reduction Scheme for OFDM Systems via Neural Networks}, 
  year={2014},
  volume={18},
  number={2},
  pages={225-228},
  doi={10.1109/LCOMM.2013.123113.131888}}

@INPROCEEDINGS{273950,
  author={Schaffer, J.D. and Whitley, D. and Eshelman, L.J.},
  booktitle={[Proceedings] COGANN-92: International Workshop on Combinations of Genetic Algorithms and Neural Networks}, 
  title={Combinations of genetic algorithms and neural networks: a survey of the state of the art}, 
  year={1992},
  volume={},
  number={},
  pages={1-37},
  doi={10.1109/COGANN.1992.273950}}

@misc{anatel_resolucao_2021,
  title = {Resolução Nº 190},
  author = {{Anatel}},
  year = {2021},
  url = {https://pesquisa.in.gov.br/imprensa/jsp/visualiza/index.jsp?data=06/10/2021\&jornal=515\&pagina=25\&totalArquivos=181}
}

@ARTICLE{7010967,
  author={Chang, Chih-Hung},
  journal={IEEE Transactions on Neural Networks and Learning Systems}, 
  title={Deep and Shallow Architecture of Multilayer Neural Networks}, 
  year={2015},
  volume={26},
  number={10},
  pages={2477-2486},
  doi={10.1109/TNNLS.2014.2387439}}

@ARTICLE{6961064,
  author={Zhao, Yongping and Wang, Kangkang},
  journal={Journal of Systems Engineering and Electronics}, 
  title={Fast cross validation for regularized extreme learning machine}, 
  year={2014},
  volume={25},
  number={5},
  pages={895-900},
  doi={10.1109/JSEE.2014.000103}}

@ARTICLE{9167435,
  author={Qi, Jun and Du, Jun and Siniscalchi, Sabato Marco and Ma, Xiaoli and Lee, Chin-Hui},
  journal={IEEE Signal Processing Letters}, 
  title={On Mean Absolute Error for Deep Neural Network Based Vector-to-Vector Regression}, 
  year={2020},
  volume={27},
  number={},
  pages={1485-1489},
  doi={10.1109/LSP.2020.3016837}}

@mastersthesis{silva2025,
  author       = {Silva, Bianca Sabrina de Cássia da},
  title        = {Técnicas de redução PAPR em sistemas de comunicações móveis com base em aprendizado de padrões},
  school       = {Instituto Nacional de Telecomunicações - INATEL},
  year         = {2025},
  address      = {Santa Rita do Sapucaí - MG, Brasil},
  type         = {Dissertação de Mestrado},
  note         = {Orientador: Prof. Dr. Luciano Leonel Mendes},
}

@INPROCEEDINGS{7899245,
  author={Khalil, Mohsin and Qadir, Junaid and Onireti, Oluwakayode and Imran, Muhammad Ali and Younis, Shahzad},
  booktitle={2017 20th Conference on Innovations in Clouds, Internet and Networks (ICIN)}, 
  title={Feasibility, architecture and cost considerations of using TVWS for rural Internet access in 5G}, 
  year={2017},
  volume={},
  number={},
  pages={23-30},
  doi={10.1109/ICIN.2017.7899245}}

@INPROCEEDINGS{10639125,
  author={da Silva, Bianca S.de C. and de Mello, Mariana B. and Mendes, Luciano L.},
  booktitle={2024 19th International Symposium on Wireless Communication Systems (ISWCS)}, 
  title={PAPR Reduction Technique for Mobile Communications System Based on Pattern Learning}, 
  year={2024},
  volume={},
  number={},
  pages={1-6},
  doi={10.1109/ISWCS61526.2024.10639125}}

\vspace{-15mm}

\begin{IEEEbiography}[{\includegraphics[width=1in,height=1.25in,clip,keepaspectratio]{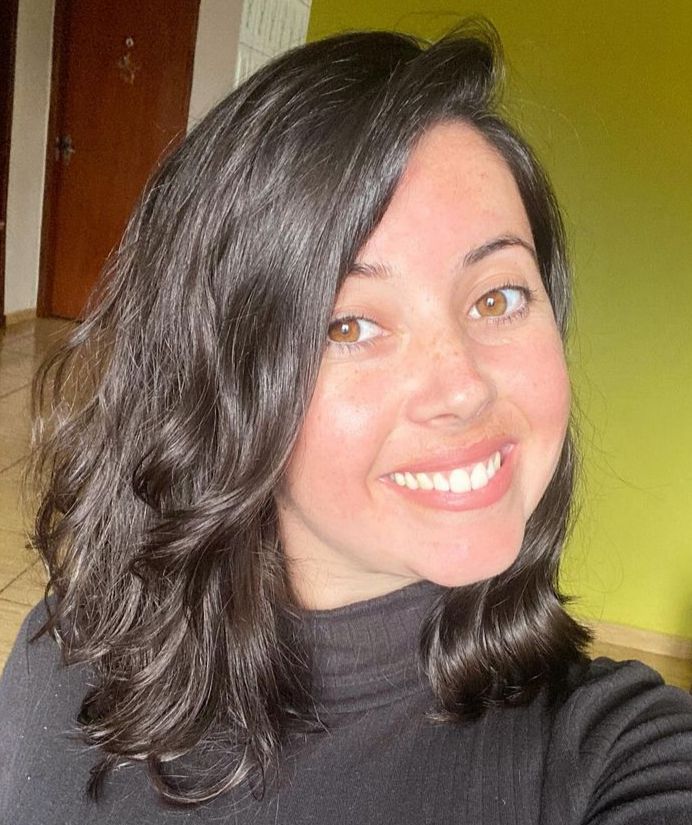}}]{Bianca S. de C. da Silva}
was born in Santa Rita do Sapucaí, Minas Gerais, Brazil, in 1998. She received the B.S. degree in Control and Automation Engineering in 2020 and the M.Sc. degree in Telecommunications Engineering in 2025, both from the National Institute of Telecommunications (INATEL), Santa Rita do Sapucaí, where she is currently pursuing a Ph.D. degree in Telecommunications Engineering. In 2023, she supported field technicians with remote site integration at Ericsson-INATEL.
\end{IEEEbiography}

\vspace{-10mm}

\begin{IEEEbiography}[{\includegraphics[width=1in,height=1.25in,clip,keepaspectratio]{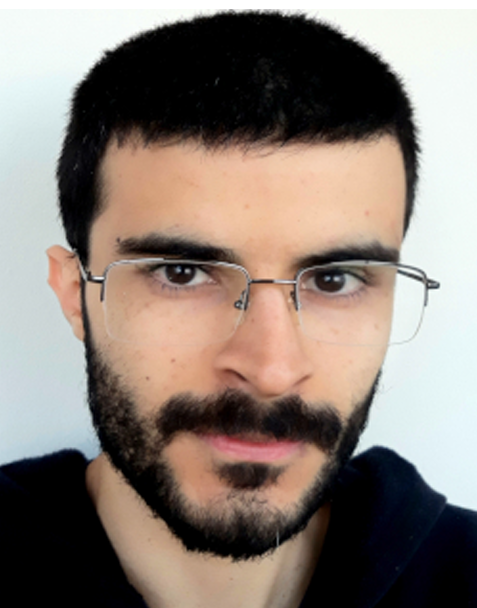}}]{Pedro H. C. de Souza}
was born in Santa Rita do Sapuca\'i, Minas Gerais, MG, Brazil in 1992. He received the B.S., M.S. and the Doctor degrees in telecommunications engineering from the National Institute of Telecommunications - INATEL, Santa Rita do Sapuca\'i, in 2015, 2017 and 2022, respectively; is currently working as a postdoctoral researcher in telecommunications engineering at INATEL, with the support of FAPESP (\textit{Fundação de Amparo à Pesquisa do Estado de São Paulo}). During the year of 2014 he was a Hardware Tester with the INATEL Competence Center - ICC. His main interests are: digital communication systems, mobile telecommunications systems, 6G, reconfigurable intelligent surfaces, convex optimization for telecommunication systems, compressive sensing/learning, cognitive radio.
\end{IEEEbiography}
\vspace{-10mm}

\begin{IEEEbiography}[{\includegraphics[width=1in,height=1.25in,clip,keepaspectratio]{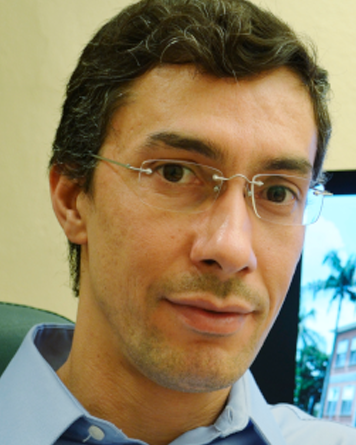}}]{Luciano L.~Mendes}
 received the B.Sc. and M.Sc. degrees from INATEL, Brazil, in 2001 and 2003, respectively, and the Ph.D. degree from Unicamp, Brazil, in 2007, all in electrical engineering. Since 2001, he has been a Professor with INATEL, where he has acted as the Technical Manager of the Hardware Development Laboratory, from 2006 to 2012. From 2013 to 2015, he was a Visiting Researcher with Vodafone Chair Mobile Communications Systems, Technical University of Dresden, where he had developed his postdoctoral training. In 2017, he was elected as the Research Coordinator of the 5G Brazil Project, an association involving industries, telecom operators, and academia, which aims for funding and build an ecosystem toward 5G in Brazil. He is the Technical Coordinator of Brazil 6G Project and general coordinator of the XGMobile - Competence Center.
\end{IEEEbiography}

\end{document}